\documentclass[prb,aps,twocolumn]{revtex4}
\usepackage{epsfig}
\begin{document}
\title{
Melting of aluminium clusters
}
\author{E. G. Noya}
\author{J. P. K. Doye} 
\affiliation{University Chemical Laboratory, Lensfield Road, Cambridge CB2 1EW,
	United Kingdom}
\author{F. Calvo}
\affiliation{Laboratoire de Physique Quantique, IRSAMC, Universit\'e Paul
Sabatier, 118 Route de Narbonne, F31062 Toulouse Cedex, France}
\date{\today}

\begin{abstract}

	The melting of Al clusters in the size range 49 $\leq N\leq$ 62
has been studied using two model interatomic potentials.
The results for the two models are significantly different. The glue
potential exhibits a smooth relatively featureless 
heat capacity curve for all sizes 
except for $N=$ 54 and $N=$ 55, sizes at
which icosahedral structures are favoured over the polytetrahedral.
Gupta heat capacity curves, instead, show a well-defined peak
that is indicative of a first-order-like transition. The differences
between the two models reflect the different ground-state structures, 
and neither potential is
able to reproduce or explain the size dependence of the
melting transition recently observed in experiments.

\end{abstract}

\maketitle
\vspace{0.5cm}
%}

\section{Introduction}

	Cluster thermodynamics has been an active field of research
during the last decades. It is now accepted that clusters 
do not melt at a singular temperature, but, instead,
there is a finite range of temperatures over which
solid-like and liquid-like isomers coexist.\cite{berry}
Nevertheless, in the cases for which a well-defined
peak is observed in the heat capacity curves, melting is
generally considered as the finite size analogue of a 
first-order phase transition.\cite{whetten}
Except for some particular cases,\cite{jarrold-1,jarrold-2} clusters 
usually melt at a temperature lower than the bulk
melting temperature, and this temperature generally decreases with the cluster 
size. The appearance of 
premelting effects, evidenced by a small peak in
the heat capacity curve or by a change of slope in the 
caloric curve before melting, is also fairly common. 

%	The cases for which experimental data is available 
%offer a unique opportunity to gain insight into 
%thermodynamic behaviour of clusters. For metal clusters, for
%example, one fundamental question is whether the 
%magic numbers in the melting temperature correlate either with
%the electronic or geometric shell closings or if, instead,
%there is an interplay between both properties. 
%
	Recently, the melting behaviour of positively
charged aluminium clusters with 49 to 62 atoms
has been measured.\cite{breaux} Except
for $N=$ 49 and 59, the heat capacity curves
show well-defined peaks between 450 and 650 K.
The melting temperature has an irregular
variation with the cluster size, suffering a sharp
drop at $N=$ 56, which was attributed to a
structural transition. A plot
of the latent heat against temperature resulted in
peaks at sizes $N=$ 51, 57 and 61. Moreover, premelting
effects were observed 100 K before melting for Al$_{51}^+$ 
and Al$_{52}^+$. 
In order to try to understand the origin
of the features observed experimentally, a theoretical study of the thermal
properties would be valuable. 

	The existing literature on 
Al clusters has mainly focused on the structural and
electronic properties. However,
in spite of the experimental\cite{pellarin,martin} and 
theoretical\cite{cheng,bernholc,jones,yang,manninen,ahlrichs,valkealahti-1,valkealahti-2,valkealahti-3,rao,johnston,lloyd-1,lloyd-2,joswig,doye,turner,lai,sebetci}
effort, a consensus has not yet been achieved about the
structure of many of the experimentally observed magic numbers.
Only for large sizes ($N > $ 250) the structure
of Al clusters seems to have been rationalized. Martin \emph{et al.}
showed that, above this size, magic numbers were due to 
geometric shell closings associated with face-centered-cubic
octahedra (fcc),\cite{martin} and this hypothesis
is consistent with results of kinetic 
simulations.\cite{valkealahti-1,valkealahti-2,valkealahti-3}
For smaller clusters, \emph{ab initio} calculations have been performed at selected
sizes.\cite{cheng,bernholc,yang,manninen,ahlrichs} 
Only at $N=$ 13 was agreement on the structure reached, which is thought to be 
icosahedral.\cite{bernholc,yang,manninen,ahlrichs} The structures at $N=$ 55 and 147 remain unassigned, as 
different theoretical calculations found very different structures. 
In particular, for $N=$ 55, icosahedral,\cite{yang} decahedral,
\cite{ahlrichs} cuboctahedral\cite{cheng} and
disordered\cite{bernholc} structures were found most stable, depending on the
method used.

	The lower computational expense of empirical potentials
have allowed more extensive searchs.\cite{lloyd-1,lloyd-2,joswig,doye,turner,lai,sebetci}
However, the results show a strong dependence on
the model potential used. The Murrell-Mottram potential\cite{mm} predicts
a competition between fcc and icosahedral structures in the size
regime $N=$ 2--55, with strong magic numbers for the 38-atom truncated
octahedron and the 54-atom uncentered Mackay icosahedron.\cite{johnston,lloyd-1,lloyd-2}
Similar results were obtained with 
Voter-Chen\cite{vc} and Gupta\cite{gupta} models. Both show
special stabilities also for the 38-atom truncated octahedron and for
the Mackay icosahedron\cite{lai,sebetci}, except that the Gupta 
potential favours the complete 
55-atom icosahedron.\cite{lai}
Very different structures were found, however, with 
the Sutton-Chen\cite{sc} and glue\cite{glue} potentials.
The former favours unusual, somewhat disordered structures, that are a
hybrid of close packed structures, decahedra and Mackay icosahedra.\cite{joswig,jon,note1}
A still different set of structures
were found with the glue model, which favours polytetrahedral
structures in the whole size range, except for those sizes close
to the complete Mackay icosahedra.\cite{doye} 

	It is the aim of the present paper to use simulation
techniques to study the melting of Al clusters in the size range
studied experimentally,\cite{breaux} in particular to see if the
experimental trends can be captured by semi-empirical potentials.
Besides, reliably finding the global minima and performing long
enough simulations to achieve well-converged thermodynamic results 
is only tractable for such potentials.
Several potentials that were fitted to Al properties have been
suggested.\cite{sc,mm,gupta,glue,vc} Of these, we have chosen 
to use the Gupta\cite{gupta} and
glue\cite{glue} potentials, as they cover two very
different set of structures.\cite{turner,lai,doye} For sodium clusters,
it has now been shown that the size dependence of the melting
behaviour primarly reflects the geometric structure of the solid
clusters.\cite{haberland} So, it is our hope that the behaviour exhibited
by these two potentials will be representative of those that
favour icosahedral and polytetrahedral structures, respectively,
in this size range. It is noteworthy that high symmetry polytetrahedral
structures are possible 
($D_{6d}$ at $N=$ 51, $D_{3h}$ at
$N=$ 57, and $T_{d}$ at $N=$ 61)\cite{morse,blj}
at the sizes for which the experimental
latent heat showed a maximum, making it particularly interesting
to study a potential that favours polytetrahedral structures.

\section{Methods}

	We performed canonical Monte Carlo (MC) simulations using 
the parallel tempering (PT) method\cite{frenkel} to study the melting
of Al clusters in the size range $N=$ 49 -- 62. 
Our simulations consisted of
10 million MC steps, following an initial equilibration period
of 1 million steps, for each of the 48 trajectories used,
whose temperatures ranged from 10 to 1000 K depending on the
cluster size and on the model. All the trajectories were 
initialized with the ground-state structures for each size and
model.\cite{lai,doye} The lowest energy geometries for
sizes not reported previously were obtained using a 
basin-hopping global optimization method.\cite{wales}
Exchange among different temperatures
was attempted with a probability of 10\%. Evaporation or
fragmentation of clusters was avoided by adding 
a spherical repulsive hard wall of radius $r_0+r$, where $r$ is the
cluster ground-state radius (calculated as the distance
from the center of mass to the furthest atom), and $r_0$ 
the equilibrium bulk interatomic distance for Gupta and 
3/2 of the dimer bond length for the glue model. These constraining
radii were large enough not to affect the location of the
main peak in the heat capacity curves. The data from these 
simulations was processed using the 
multihistogram technique.\cite{whetten}

\begin{figure}
\begin{center}
\includegraphics[width=86mm]{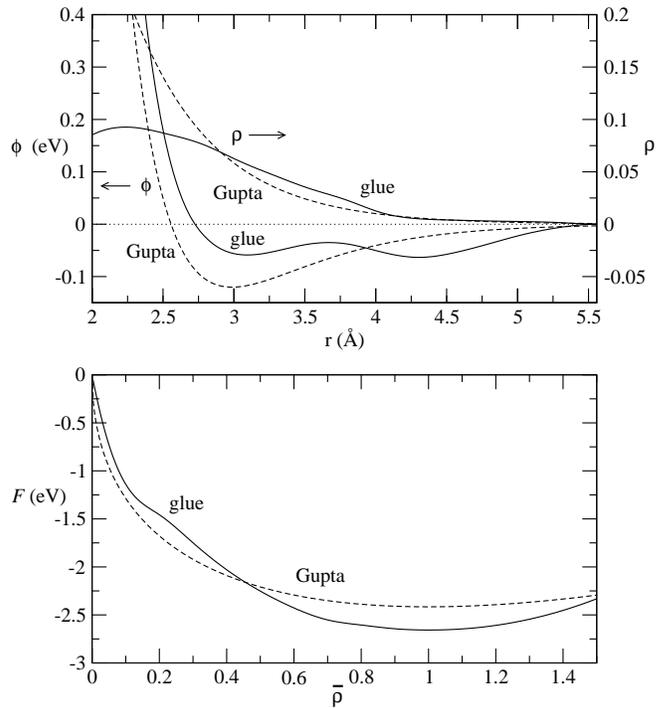}
\caption{\label{pots} Comparison of $\phi (r)$, $\rho (r)$ and $F(\bar{\rho })$ for
Al Gupta and glue potentials in the effective pair format.}
\end{center}
\end{figure}

	The two potentials used belong to the family of embedded-atom potentials.
They comprise a pair potential plus a many-body term that
depends on the electronic density at the atomic position. Therefore, the
total energy can be written:
\begin{equation}
 E = \sum_{i} \sum_{j\neq i} \phi ( r_{ij} ) +
 	\sum_{i} F (\bar{\rho }_i ) ,
\end{equation}
where $\phi (r_{ij})$ is a pair potential, $r_{ij}$ is the
interatomic distance between atoms $i$ and $j$, $F(\bar{\rho}_i)$
is an embedding function, and $\bar{\rho }_i$ is
the electronic density at
site $i$, which is usually approximated as the linear superposition
of the atomic densities of the rest of the atoms, 
i.e.\ $\bar{\rho }_i = \sum_{j\neq i} \rho(r_{ij})$.
The main difference between the two potentials is that
Gupta assumes analytic forms for the pair potential, the embedding
function and the density,\cite{gupta} 
whereas, in the glue model, these functions are determined
by the fitting process.\cite{glue} The best way to compare
the two potentials is in the effective pair format.\cite{johnson} The forms
for $\phi (r)$ and $F(\bar{\rho })$ are non-unique, and in the 
effective pair format they are chosen so that the embedding
function has a minimum at the value of $\bar{\rho }$ appropriate
for the crystal. As can be seen from Fig. \ref{pots}, there are
considerable differences between the two potentials.
For example, the pair potential for the glue potential
is significantly shallower,
and shows a double well structure as compared to the 
single well in the Gupta pair potential. 

	The potentials also differ in the number 
of properties and configurations that have
been used in the fitting process.
The glue model was adjusted to reproduce the forces obtained by
first-principles calculations for a variety of environments, including
surfaces, clusters, liquids and
crystals.\cite{glue} However, the four free parameters in the
Gupta potential have been simply fitted to 
some Al bulk properties, namely, 
the lattice parameters and elastic moduli.\cite{cleri}
Therefore, the glue potential is more likely to work well
in a variety of situations, and, for example, has been
particularly successful in modelling the self-diffusion
in Al.\cite{sandberg}

\begin{figure*}
\begin{center}
\includegraphics[width=180mm]{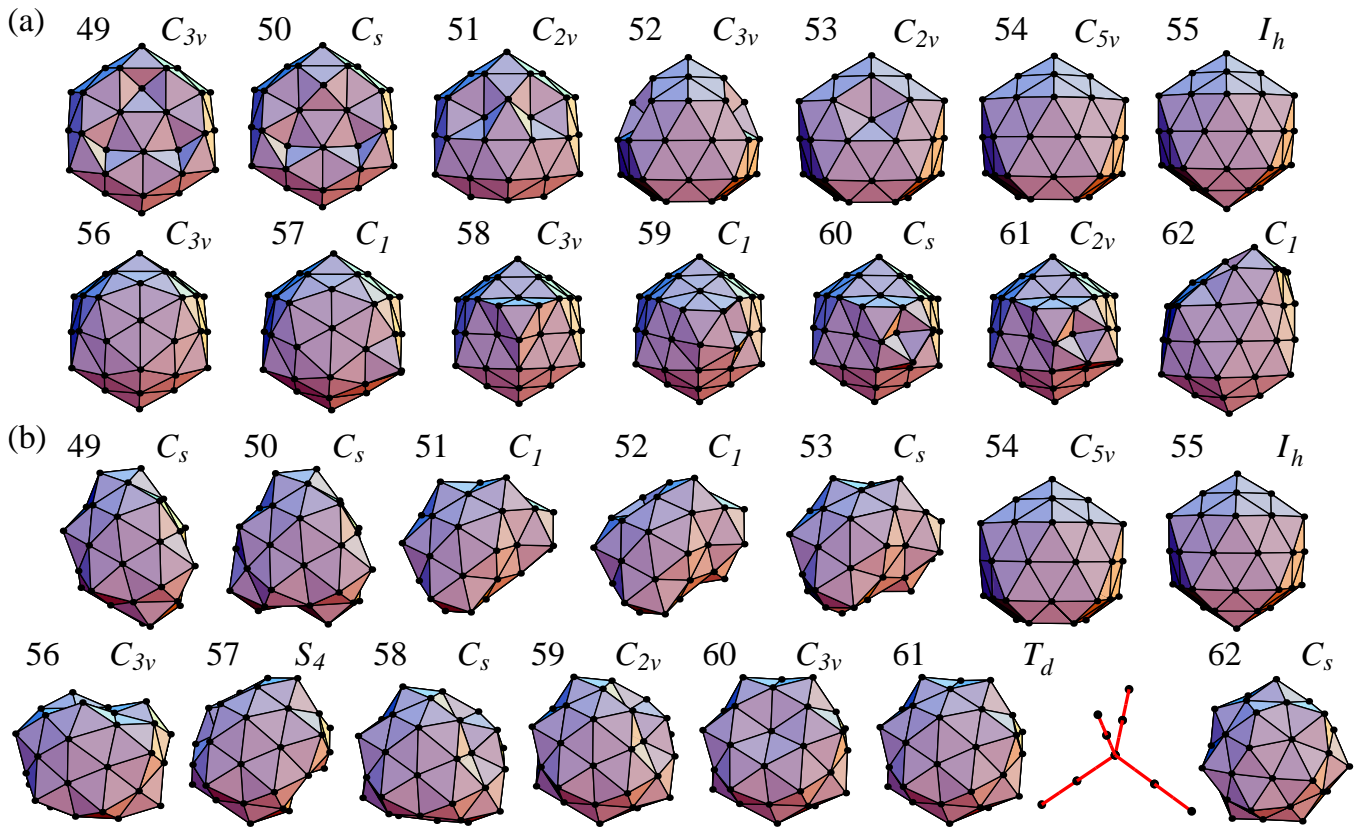}
\caption{\label{figmin} (Colour online) Ground state structures for Al (a) 
Gupta and (b) glue 
clusters with 49 to 62 atoms. For the 61-atom glue cluster,
the disclination network is also depicted.}
\end{center}
\end{figure*}

	For both potentials some thermodynamic properties of the
Al clusters have been studied previously,\cite{jellinek,lai,werner,sun}
but the only overlap with the size range considered here are
at $N=$ 54--56 for the Gupta potential.\cite{jellinek,lai,werner}

\section{Results}

	The structures of the global minima for this size range are 
depicted in Fig.\ref{figmin}. For the Gupta potential, all the 
structures are based upon the Mackay icosahedron, whereas for the
glue potential, they are all polytetrahedral (i.e., the whole
structure can be divided up into tetrahedra with atoms at the
vertices), except at
$N=$ 54 and 55, which are icosahedral. For the Gupta potential,
the 55-atom Mackay icosahedron is most stable, whereas
for the glue potential, the 55-atom and 61-atom structures are 
particularly stable. For polytetrahedral
structures, it is generally preferred to have five tetrahedra
around a nearest-neighbour contact, but beyond a certain size, edges 
surrounded by six tetrahedra must also be present.
The network formed by these sixfold edges, termed 
disclination network, provides an useful way to characterize 
polytetrahedral structures. For $N=$ 61 the tetrahedral disclination network
is depicted in Fig. \ref{figmin}, and most of the other
polytetrahedral structures are based upon this structure. It is noteworthy
that the possible high-symmetry polytetrahedral structures
at $N=$ 51 and 57, which involve a linear and a trigonal
disclination network, mentioned in the introduction as potential candidates for the
experimental peaks in the latent heat, are not favoured by
this potential. However, whether the 61-atom structure leads
to a particularly large latent heat, should provide an indicator
of whether this suggestion could be correct.

	That the two potentials show completely different structures
does not necessarily mean that one or both are bad potentials, but
illustrates how difficult it is for a potential
to correctly predict a cluster's structure, because to do so, the
potential must be able to model a whole host of
bulk and surface properties of the material correctly. Indeed, it is
not uncommon for potentials that purport to model the same
material to exhibit very different structures.\cite{lead}

\begin{figure}
\begin{center}
\includegraphics[width=86mm]{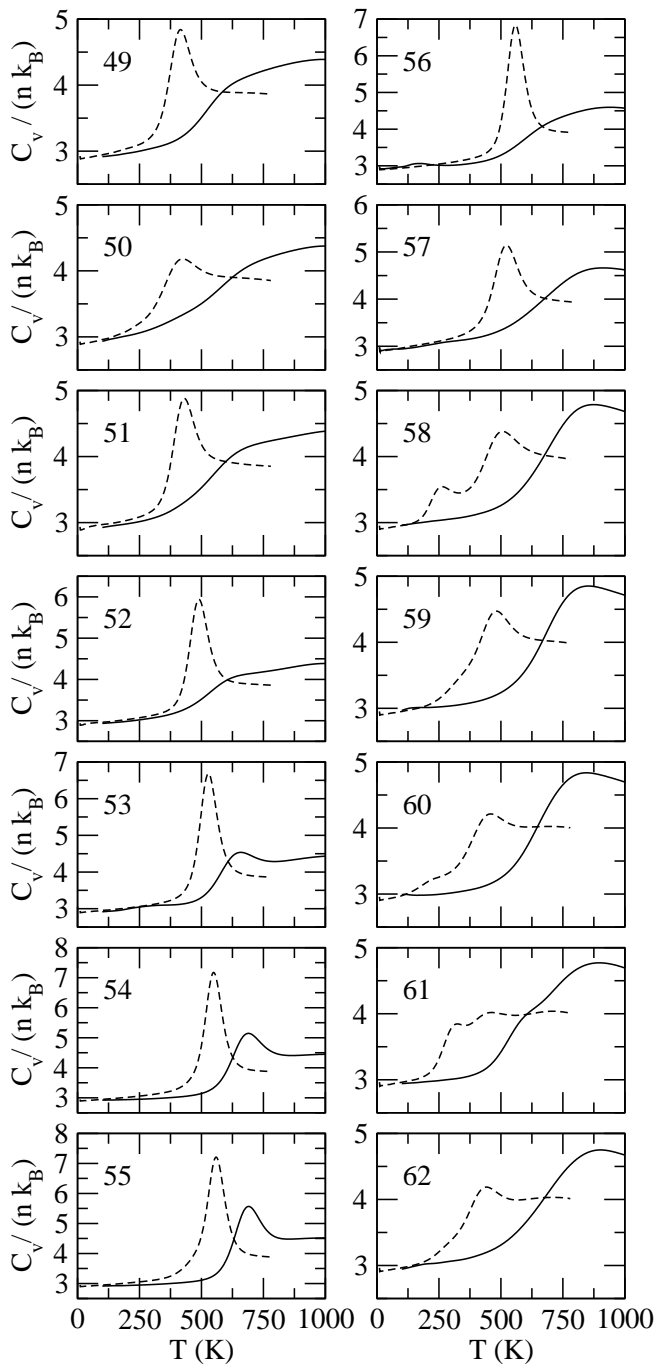}
\caption{\label{fig1} Heat capacities as a function of temperature for
aluminium clusters from 49 to 62 atoms, predicted by the glue 
(solid line) and Gupta (dashed line) models.}
\end{center}
\end{figure}

	Figure \ref{fig1} shows the calculated canonical heat 
capacities as a function of temperature
for Al clusters with 49 to 62 atoms using the Gupta and
glue models. The heat capacity curves for the two potentials
show few similarities. The Gupta heat capacities show a fairly 
well-defined peak for almost all sizes, as expected for clusters
with icosahedral geometries, and the results for $N=$ 54--56 are
consistent with previous results.\cite{jellinek,werner,lai}
Interestingly, premelting effects are observed for the sizes $N=$ 58--62. 

	The glue potential, instead, predicts a smooth
transition from the solid to the liquid state without an
associated latent heat or peak in the heat capacity curve.
The sizes $N=$ 54 and 55,
which have icosahedral ground-state structures, represent an exception,
and their heat capacity curves are more similar to that for the
Gupta potential, showing a well-defined peak albeit broader
and at higher temperature than for Gupta. Similar behaviour was
observed in a previous simulation study using the glue model.\cite{sun}
Sun and Gong found a well-defined peak in the heat
capacity of the icosahedral clusters at $N=$ 13 and 147, but
the polytetrahedral 43-atom cluster underwent a more continuous transition,
without a clear peak in the heat capacity.
Small peaks are found in the low
temperature region of the heat capacity curves of the clusters
Al$_{53}$, Al$_{56}$ and Al$_{57}$,
which are indicative of premelting effects.

\begin{figure}
\begin{center}
\includegraphics[width=85mm]{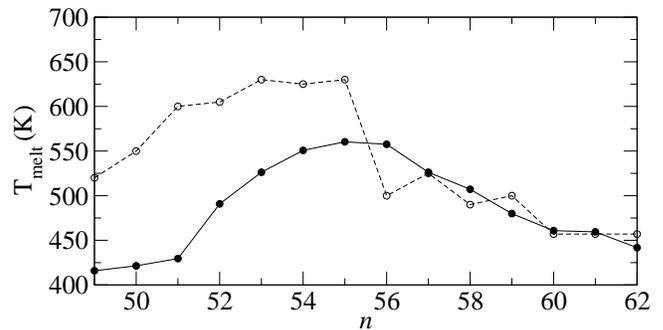}
\caption{\label{fig2} Comparison of the experimental
melting temperatures (open circles) with those predicted
by Gupta (closed circles).}
\end{center}
\end{figure}

	The fact that the two potentials predict significantly 
different melting behaviours is not surprising, given the differences
in the structure of the lowest-energy clusters. The absence of a 
well-defined heat capacity peak for clusters with polytetrahedral
global minima reflects the basic structural similarities between
the ground-state and the molten clusters. It is well-known
that simple liquids have substantial polytetrahedral character.\cite{nelson}
Hence, melting in these clusters is associated with the gradual
occupation of structurally similar isomers of higher and
higher energy, rather than the more usual cooperative transition
between two sets of structures that have different energies 
and entropies. Similar behaviour is also seen for Lennard-Jones
clusters with 25--30 atoms, as they have polytetrahedral
global minima.\cite{frantz}
A more cooperative, first-order-like transition only appears
once the structure of the global minimum changes. Sun and Gong 
were right to describe such a continuous melting as more akin to 
the melting of a ``glass" (or perhaps more properly an ``ideal glass"
as the clusters are always in equilibrium) than the melting of an
ordered solid that has a fundamentally different `symmetry' to 
that of the liquid.

\begin{figure}
\begin{center}
\includegraphics[width=85mm]{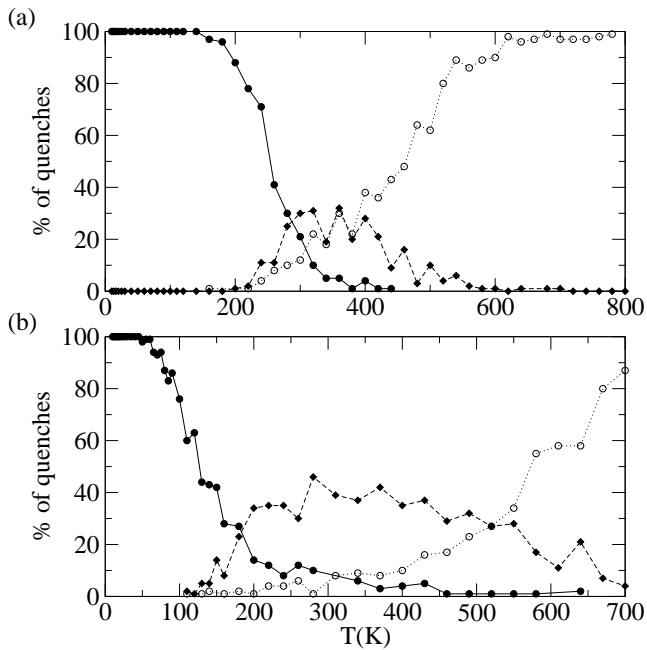}
\caption{\label{fig3} A structural
analysis of the melting of the (a) 
Gupta Al$_{58}$ and (b) glue Al$_{56}$ clusters.
The percentage of
quenches leading to the icosahedral ground-state
structure (closed circles), to structures which have absorbed some
or all the extra 
atoms in the surface (diamonds), and to high energy
isomers (open circles) is plotted. 
In (b) the structures considered are
the polytetrahedral ground-state structure (closed circles), 
an icosahedral isomer (diamonds) and high
energy isomers (open circles).}
\end{center}
\end{figure}

	The results obtained by either potential differ significantly from
those obtained in the experiments of Breaux \emph{et al.}\cite{breaux} 
Of the two models studied, only the Gupta heat capacities show
well-defined peaks for most sizes, as observed
in experiments. The continuous nature of the melting transition
for most of the glue Al clusters qualitatively disagrees
with experiments. Breaux \emph{et al.} also reported
a non-monotonic variation of the melting temperature 
with the cluster size, with a maximum at $N=$ 55 and a
sharp drop for larger sizes. 
In Fig. \ref{fig2}, the size dependence of the Gupta and
experimental melting temperatures is compared.
There is some agreement, but also significant differences. As one
would expect for clusters with Mackay-based structures,
the Gupta clusters exhibit a maximum at the complete 55-Mackay icosahedron
magic number,
and the melting temperature decreases monotonically on either adding
or removing atoms. However, this maximum is broad and there is not
the sharp decrease seen experimentally in going from $N=$ 55 to 56. 
This broadness is probably due to compensating changes to both
the energy and entropy of melting.\cite{schmidt} Instead, the latent heat
of melting is a more structurally sensitive quantity.\cite{haberland}
The latent heats obtained from the Gupta clusters show a
size dependence very similar to the one found for the melting temperature (not shown),
and the experimentally observed peaks at $N=$ 51, 57 and 61 
are not reproduced by our simulations.
As for the sizes at which premelting effects occur,
there is disagreement both between our results and the
experiments, and between the two models. In particular,
the premelting effects at $N=$ 51 and 52 were not reproduced by
either of the models.

	Even though experiments concern positively charged clusters 
and our calculations were performed for neutral clusters, it is unlikely
that the discrepancies can be attributed to the charge.
The effect that charge might have in such large clusters
is probably small, as has been illustrated for Na clusters.\cite{calvo}
Rather, deficiencies in the potentials, and hence the 
structures they predict, are the more likely cause.

	In what follows we will analyse the origin of 
the premelting effects for both potentials. For that purpose,
the microscopic behaviour of the system was followed by performing
systematic quenches starting from each one of the trajectories.
The Al$_{58}$ Gupta and Al$_{56}$ glue clusters were chosen
as examples that show premelting effects with each
potential. 

	In Fig. \ref{fig3} we have plotted the percentage of quenches
that lead to a particular minimum for Gupta Al$_{58}$.
The ground-state structure of this cluster is
a Mackay icosahedron, with the three adatoms located at anti-Mackay
positions, all on the same face (Fig.\ \ref{figmin}).
The percentage of quenches that go to this structure is very high at low 
temperatures, but
drops abruptly between 200 and 300 K. Between approximately
250 K to 500 K, the lowest-energy
minimum coexists with a family of isomers in which some or all the extra 
atoms, that lie above the surface in the ground state, are accommodated
in the surface. Note that similar structures consisting of 
a Mackay icosahedron with one of the extra atoms absorbed in its surface are the
ground-state structures for Al$_{56}$ and Al$_{57}$ (Fig.\ \ref{figmin}). 
At temperatures higher than 550 K, high energy isomers dominate and
the cluster is completely melted. 
Therefore, the first peak in the heat capacity of Gupta Al$_{58}$
can be attributed to a surface reconstruction in which some or all the extra
atoms accommodate in the surface. The same type of surface
reconstruction was seen for the other Gupta clusters that showed 
premelting effects, e.g.\ for Al$_{57}$
an isomer in which the two extra atoms are accommodated in the surface
becomes competitive above 200 K.
Intuitively, these structures will become less favourable as the number
of atoms in the third shell of the Mackay icosahedron increases, which is consistent with the
fact that premelting features become more subtle for larger sizes.

%Even though this situation 
%increases the strain in the surface, it is also accompanied by a higher surface
%energy. This kind of structures are expected to be favourable
%for systems that are able to accommodate a large strain without
%a large energy penalty. Due to the
%broad minima in the embedding function of metallic models, which, besides,
%only depends on the density and not on the relative distances,
%the strained structures can become very competitive. 

	For all sizes at which the glue model exhibits premelting 
effects, namely, $N=$ 53, 56 and 57, 
the lowest-lying minima are polytetrahedral.\cite{doye}
However, for Al$_{56}$
an icosahedral isomer becomes competitive at 180 K, 
and approximately 50\% of the quenches 
lead to this minimum up to approximately
580 K (see Fig. \ref{fig3}). Only above this 
temperature do high-energy isomers dominate.
Similar analyses with analogous results were also performed
for Al$_{53}$ and Al$_{56}$. From the microscopic analysis
it seems clear that the 
premelting effects observed in the glue model
can be attributed to this solid-solid 
transition from polytetrahedral to icosahedral structures.

	Such a transition from polytetrahedral
to icosahedral structures 
is somehow surprising. Even
though icosahedral structures are energetically competitive 
for the nearest sizes to a shell closing,\cite{doye}
the solid-solid transition can only be explained if
the icosahedral isomers have
higher vibrational entropy. However, for pair potentials
at least, this is not the usual situation; instead,
polytetrahedral structures usually have a higher
vibrational entropy because of the greater internal strains.\cite{doye-4}
The situation is reversed here.
For Al$_{56}$, the geometric mean vibrational frequency
is 1.0533 times larger for the polytetrahedral
structure, which means that they have a lower vibrational entropy.
An estimate of the solid-solid transition can hence be obtained 
within the harmonic approximation.\cite{doye-4} Following
this procedure, the solid-solid transition is predicted
to take place at 160 K, which is in very good agreement with the 
temperature of 168 K obtained from our PT simulations.

\section{Conclusions}

	We have studied the melting behaviour of Al
clusters in the range $N=$ 49--62 by means of PT MC simulations
using two different models.
The results are qualitatively and quantitatively different depending
on the potential used, and neither of the potentials is able to reproduce the
melting behaviour observed experimentally.\cite{breaux}
Many of the differences in the melting behaviour of the two models 
can be attributed to the different lowest-energy structures, which are mainly
polytetrahedral for the glue potential and predominantly icosahedral
for the Gupta potential. These results therefore suggest that neither
polytetrahedral or icosahedral structures can explain the
experimental size-dependence of the melting behaviour. Instead,
the actual structure of Al clusters in this size range 
remains a mystery. Furthermore, we feel that none of the
other semiempirical potentials are any more likely to
be successful in reproducing the Al thermal properties.
An explanation of the experimental results would need, therefore, the application
of models at a higher level of theory, but the computational
demands of those techniques are significantly higher.

	Nevertheless, the results shown here have helped to understand
what kind of melting behaviour can be expected from
potentials with different structural preferences. The melting 
transition for potentials that favour polytetrahedral structures
is more akin to an ideal glass transition than the finite size
equivalent of the bulk first-order melting transition.
It is particularly interesting that this more continuous
character of the melting transition even holds for
sizes, such as $N=$ 61, at which the polytetrahedral
global minimum is particularly stable.
Also, due to the larger relative importance of
the many-body term over the pair potential, unusual features, 
such as the solid-solid transition
from polytetrahedral to icosahedral structure driven by
the vibrational entropy, can be found. However, models that exhibit
icosahedral structures will tend to exhibit well-defined
heat capacity peaks akin to bulk melting.

%\vspace*{5cm}
%$^{\rm a)}$Author to whom correspondence should be addressed. Electronic mail:
%\mbox{}

%\narrowtext
	
%\footnote[1]{As $n/m$ value for the Al SC potential is close
%to one in the effective pair format, the potential has a shallow well
%in the pair potential and so the observed structures are somewhat
%similar to those seen for Zn and Cd Gupta potentials, which also have
%very shallow wells in their effective pair potentials.}

\acknowledgements
	The authors are grateful to the Ram\'on Areces Foundation (E.G.N.) 
and the Royal Society (J.P.K.D.) for financial support.

%\bibliography{aluminium}

\begin{thebibliography}{50}
\expandafter\ifx\csname natexlab\endcsname\relax\def\natexlab#1{#1}\fi
\expandafter\ifx\csname bibnamefont\endcsname\relax
  \def\bibnamefont#1{#1}\fi
\expandafter\ifx\csname bibfnamefont\endcsname\relax
  \def\bibfnamefont#1{#1}\fi
\expandafter\ifx\csname citenamefont\endcsname\relax
  \def\citenamefont#1{#1}\fi
\expandafter\ifx\csname url\endcsname\relax
  \def\url#1{\texttt{#1}}\fi
\expandafter\ifx\csname urlprefix\endcsname\relax\def\urlprefix{URL }\fi
\providecommand{\bibinfo}[2]{#2}
\providecommand{\eprint}[2][]{\url{#2}}

\bibitem[{\citenamefont{Berry et~al.}(1984)\citenamefont{Berry, Jellinek, and
  Natanson}}]{berry}
\bibinfo{author}{\bibfnamefont{R.~S.} \bibnamefont{Berry}},
  \bibinfo{author}{\bibfnamefont{J.}~\bibnamefont{Jellinek}}, \bibnamefont{and}
  \bibinfo{author}{\bibfnamefont{G.}~\bibnamefont{Natanson}},
  \bibinfo{journal}{Chem. Phys. Lett.} \textbf{\bibinfo{volume}{107}},
  \bibinfo{pages}{227} (\bibinfo{year}{1984}).

\bibitem[{\citenamefont{Labastie and Whetten}(1990)}]{whetten}
\bibinfo{author}{\bibfnamefont{P.}~\bibnamefont{Labastie}} \bibnamefont{and}
  \bibinfo{author}{\bibfnamefont{R.~L.} \bibnamefont{Whetten}},
  \bibinfo{journal}{Phys. Rev. Lett.} \textbf{\bibinfo{volume}{65}},
  \bibinfo{pages}{1567} (\bibinfo{year}{1990}).

\bibitem[{\citenamefont{Shvartsburg and Jarrold}(2000)}]{jarrold-1}
\bibinfo{author}{\bibfnamefont{A.~A.} \bibnamefont{Shvartsburg}}
  \bibnamefont{and} \bibinfo{author}{\bibfnamefont{M.~F.}
  \bibnamefont{Jarrold}}, \bibinfo{journal}{Phys. Rev. Lett.}
  \textbf{\bibinfo{volume}{85}}, \bibinfo{pages}{2530} (\bibinfo{year}{2000}).

\bibitem[{\citenamefont{Breaux et~al.}(2003)\citenamefont{Breaux, Benirschke,
  Sugai, Kinnear, and Jarrold}}]{jarrold-2}
\bibinfo{author}{\bibfnamefont{G.~A.} \bibnamefont{Breaux}},
  \bibinfo{author}{\bibfnamefont{R.~C.} \bibnamefont{Benirschke}},
  \bibinfo{author}{\bibfnamefont{T.}~\bibnamefont{Sugai}},
  \bibinfo{author}{\bibfnamefont{B.~S.} \bibnamefont{Kinnear}},
  \bibnamefont{and} \bibinfo{author}{\bibfnamefont{M.~F.}
  \bibnamefont{Jarrold}}, \bibinfo{journal}{Phys. Rev. Lett.}
  \textbf{\bibinfo{volume}{91}}, \bibinfo{pages}{215508}
  (\bibinfo{year}{2003}).

\bibitem[{\citenamefont{Breaux et~al.}(2005)\citenamefont{Breaux, Neal, Cao,
  and Jarrold}}]{breaux}
\bibinfo{author}{\bibfnamefont{G.~A.} \bibnamefont{Breaux}},
  \bibinfo{author}{\bibfnamefont{C.~M.} \bibnamefont{Neal}},
  \bibinfo{author}{\bibfnamefont{B.}~\bibnamefont{Cao}}, \bibnamefont{and}
  \bibinfo{author}{\bibfnamefont{M.~F.} \bibnamefont{Jarrold}},
  \bibinfo{journal}{Phys. Rev. Lett.} \textbf{\bibinfo{volume}{94}},
  \bibinfo{pages}{173401} (\bibinfo{year}{2005}).

\bibitem[{\citenamefont{Pellarin et~al.}(1993)\citenamefont{Pellarin,
  Baguenard, Broyer, Lerm\'{e}, and Vialle}}]{pellarin}
\bibinfo{author}{\bibfnamefont{M.}~\bibnamefont{Pellarin}},
  \bibinfo{author}{\bibfnamefont{B.}~\bibnamefont{Baguenard}},
  \bibinfo{author}{\bibfnamefont{M.}~\bibnamefont{Broyer}},
  \bibinfo{author}{\bibfnamefont{J.}~\bibnamefont{Lerm\'{e}}},
  \bibnamefont{and} \bibinfo{author}{\bibfnamefont{J.~L.}
  \bibnamefont{Vialle}}, \bibinfo{journal}{J. Chem. Phys.}
  \textbf{\bibinfo{volume}{98}}, \bibinfo{pages}{944} (\bibinfo{year}{1993}).

\bibitem[{\citenamefont{Martin et~al.}(1992)\citenamefont{Martin, N{\"a}her,
  and Schaber}}]{martin}
\bibinfo{author}{\bibfnamefont{T.~P.} \bibnamefont{Martin}},
  \bibinfo{author}{\bibfnamefont{U.}~\bibnamefont{N{\"a}her}},
  \bibnamefont{and} \bibinfo{author}{\bibfnamefont{H.}~\bibnamefont{Schaber}},
  \bibinfo{journal}{Chem. Phys. Lett.} \textbf{\bibinfo{volume}{199}},
  \bibinfo{pages}{470} (\bibinfo{year}{1992}).

\bibitem[{\citenamefont{Cheng et~al.}(1991)\citenamefont{Cheng, Berry, and
  Whetten}}]{cheng}
\bibinfo{author}{\bibfnamefont{H.-P.} \bibnamefont{Cheng}},
  \bibinfo{author}{\bibfnamefont{R.~S.} \bibnamefont{Berry}}, \bibnamefont{and}
  \bibinfo{author}{\bibfnamefont{R.~L.} \bibnamefont{Whetten}},
  \bibinfo{journal}{Phys. Rev. B} \textbf{\bibinfo{volume}{43}},
  \bibinfo{pages}{10647} (\bibinfo{year}{1991}).

\bibitem[{\citenamefont{Yi et~al.}(1991)\citenamefont{Yi, Oh, and
  Bernholc}}]{bernholc}
\bibinfo{author}{\bibfnamefont{J.-Y.} \bibnamefont{Yi}},
  \bibinfo{author}{\bibfnamefont{D.~J.} \bibnamefont{Oh}}, \bibnamefont{and}
  \bibinfo{author}{\bibfnamefont{J.}~\bibnamefont{Bernholc}},
  \bibinfo{journal}{Phys. Rev. Lett.} \textbf{\bibinfo{volume}{67}},
  \bibinfo{pages}{1549} (\bibinfo{year}{1991}).

\bibitem[{\citenamefont{Jones}(1991)}]{jones}
\bibinfo{author}{\bibfnamefont{R.~O.} \bibnamefont{Jones}},
  \bibinfo{journal}{Phys. Rev. Lett.} \textbf{\bibinfo{volume}{67}},
  \bibinfo{pages}{224} (\bibinfo{year}{1991}).

\bibitem[{\citenamefont{Yang et~al.}(1993)\citenamefont{Yang, Drabold, Adams,
  and Sachdev}}]{yang}
\bibinfo{author}{\bibfnamefont{S.~H.} \bibnamefont{Yang}},
  \bibinfo{author}{\bibfnamefont{D.~A.} \bibnamefont{Drabold}},
  \bibinfo{author}{\bibfnamefont{J.~B.} \bibnamefont{Adams}}, \bibnamefont{and}
  \bibinfo{author}{\bibfnamefont{A.}~\bibnamefont{Sachdev}},
  \bibinfo{journal}{Phys. Rev. B} \textbf{\bibinfo{volume}{47}},
  \bibinfo{pages}{1567} (\bibinfo{year}{1993}).

\bibitem[{\citenamefont{Akola et~al.}(1999)\citenamefont{Akola, Manninen,
  H{\"a}kkinen, Landman, Li, and Wang}}]{manninen}
\bibinfo{author}{\bibfnamefont{J.}~\bibnamefont{Akola}},
  \bibinfo{author}{\bibfnamefont{M.}~\bibnamefont{Manninen}},
  \bibinfo{author}{\bibfnamefont{H.}~\bibnamefont{H{\"a}kkinen}},
  \bibinfo{author}{\bibfnamefont{U.}~\bibnamefont{Landman}},
  \bibinfo{author}{\bibfnamefont{X.}~\bibnamefont{Li}}, \bibnamefont{and}
  \bibinfo{author}{\bibfnamefont{L.-S.} \bibnamefont{Wang}},
  \bibinfo{journal}{Phys. Rev. B} \textbf{\bibinfo{volume}{60}},
  \bibinfo{pages}{R11297} (\bibinfo{year}{1999}).

\bibitem[{\citenamefont{Ahlrichs and Elliott}(1999)}]{ahlrichs}
\bibinfo{author}{\bibfnamefont{R.}~\bibnamefont{Ahlrichs}} \bibnamefont{and}
  \bibinfo{author}{\bibfnamefont{S.~D.} \bibnamefont{Elliott}},
  \bibinfo{journal}{Phys. Chem. Chem. Phys.} \textbf{\bibinfo{volume}{1}},
  \bibinfo{pages}{13} (\bibinfo{year}{1999}).

\bibitem[{\citenamefont{Valkealahti and Manninen}(1994)}]{valkealahti-1}
\bibinfo{author}{\bibfnamefont{S.}~\bibnamefont{Valkealahti}} \bibnamefont{and}
  \bibinfo{author}{\bibfnamefont{M.}~\bibnamefont{Manninen}},
  \bibinfo{journal}{Phys. Rev. B} \textbf{\bibinfo{volume}{50}},
  \bibinfo{pages}{17564} (\bibinfo{year}{1994}).

\bibitem[{\citenamefont{Valkealahti et~al.}(1995)\citenamefont{Valkealahti,
  N{\"a}her, and Manninen}}]{valkealahti-2}
\bibinfo{author}{\bibfnamefont{S.}~\bibnamefont{Valkealahti}},
  \bibinfo{author}{\bibfnamefont{U.}~\bibnamefont{N{\"a}her}},
  \bibnamefont{and} \bibinfo{author}{\bibfnamefont{M.}~\bibnamefont{Manninen}},
  \bibinfo{journal}{Phys. Rev. B} \textbf{\bibinfo{volume}{51}},
  \bibinfo{pages}{11039} (\bibinfo{year}{1995}).

\bibitem[{\citenamefont{Valkealahti and Manninen}(1998)}]{valkealahti-3}
\bibinfo{author}{\bibfnamefont{S.}~\bibnamefont{Valkealahti}} \bibnamefont{and}
  \bibinfo{author}{\bibfnamefont{M.}~\bibnamefont{Manninen}},
  \bibinfo{journal}{Phys. Rev. B} \textbf{\bibinfo{volume}{57}},
  \bibinfo{pages}{15533} (\bibinfo{year}{1998}).

\bibitem[{\citenamefont{Rao and Jena}(1999)}]{rao}
\bibinfo{author}{\bibfnamefont{B.~K.} \bibnamefont{Rao}} \bibnamefont{and}
  \bibinfo{author}{\bibfnamefont{P.}~\bibnamefont{Jena}}, \bibinfo{journal}{J.
  Chem. Phys.} \textbf{\bibinfo{volume}{111}}, \bibinfo{pages}{1890}
  (\bibinfo{year}{1999}).

\bibitem[{\citenamefont{Johnston and Fang}(1992)}]{johnston}
\bibinfo{author}{\bibfnamefont{R.~L.} \bibnamefont{Johnston}} \bibnamefont{and}
  \bibinfo{author}{\bibfnamefont{J.-Y.} \bibnamefont{Fang}},
  \bibinfo{journal}{J. Chem. Phys.} \textbf{\bibinfo{volume}{97}},
  \bibinfo{pages}{7809} (\bibinfo{year}{1992}).

\bibitem[{\citenamefont{Lloyd and Johnston}(1998)}]{lloyd-1}
\bibinfo{author}{\bibfnamefont{L.~D.} \bibnamefont{Lloyd}} \bibnamefont{and}
  \bibinfo{author}{\bibfnamefont{R.~L.} \bibnamefont{Johnston}},
  \bibinfo{journal}{Chem. Phys.} \textbf{\bibinfo{volume}{236}},
  \bibinfo{pages}{107} (\bibinfo{year}{1998}).

\bibitem[{\citenamefont{Lloyd et~al.}(2002)\citenamefont{Lloyd, Johnston,
  Roberts, and Mortimer-Jones}}]{lloyd-2}
\bibinfo{author}{\bibfnamefont{L.~D.} \bibnamefont{Lloyd}},
  \bibinfo{author}{\bibfnamefont{R.~L.} \bibnamefont{Johnston}},
  \bibinfo{author}{\bibfnamefont{C.}~\bibnamefont{Roberts}}, \bibnamefont{and}
  \bibinfo{author}{\bibfnamefont{T.~V.} \bibnamefont{Mortimer-Jones}},
  \bibinfo{journal}{ChemPhysChem} \textbf{\bibinfo{volume}{3}},
  \bibinfo{pages}{408} (\bibinfo{year}{2002}).

\bibitem[{\citenamefont{Joswig and Springborg}(2003)}]{joswig}
\bibinfo{author}{\bibfnamefont{J.}~\bibnamefont{Joswig}} \bibnamefont{and}
  \bibinfo{author}{\bibfnamefont{M.}~\bibnamefont{Springborg}},
  \bibinfo{journal}{Phys. Rev. B} \textbf{\bibinfo{volume}{68}},
  \bibinfo{pages}{085408} (\bibinfo{year}{2003}).

\bibitem[{\citenamefont{Doye}(2003{\natexlab{a}})}]{doye}
\bibinfo{author}{\bibfnamefont{J.~P.~K.} \bibnamefont{Doye}},
  \bibinfo{journal}{J. Chem. Phys.} \textbf{\bibinfo{volume}{119}},
  \bibinfo{pages}{1136} (\bibinfo{year}{2003}{\natexlab{a}}).

\bibitem[{\citenamefont{Turner et~al.}(2000)\citenamefont{Turner, Johnston, and
  Wilson}}]{turner}
\bibinfo{author}{\bibfnamefont{G.~W.} \bibnamefont{Turner}},
  \bibinfo{author}{\bibfnamefont{R.~L.} \bibnamefont{Johnston}},
  \bibnamefont{and} \bibinfo{author}{\bibfnamefont{N.~T.}
  \bibnamefont{Wilson}}, \bibinfo{journal}{J. Chem. Phys.}
  \textbf{\bibinfo{volume}{112}}, \bibinfo{pages}{4773} (\bibinfo{year}{2000}).

\bibitem[{\citenamefont{Lai et~al.}(2004)\citenamefont{Lai, Lin, Wu, Li, and
  Lee}}]{lai}
\bibinfo{author}{\bibfnamefont{S.~K.} \bibnamefont{Lai}},
  \bibinfo{author}{\bibfnamefont{W.~D.} \bibnamefont{Lin}},
  \bibinfo{author}{\bibfnamefont{K.~L.} \bibnamefont{Wu}},
  \bibinfo{author}{\bibfnamefont{W.~H.} \bibnamefont{Li}}, \bibnamefont{and}
  \bibinfo{author}{\bibfnamefont{K.~C.} \bibnamefont{Lee}},
  \bibinfo{journal}{J. Chem. Phys.} \textbf{\bibinfo{volume}{121}},
  \bibinfo{pages}{1487} (\bibinfo{year}{2004}).

\bibitem[{\citenamefont{Sebetci and G{\"u}ven\c{c}}(2005)}]{sebetci}
\bibinfo{author}{\bibfnamefont{A.}~\bibnamefont{Sebetci}} \bibnamefont{and}
  \bibinfo{author}{\bibfnamefont{Z.~B.} \bibnamefont{G{\"u}ven\c{c}}},
  \bibinfo{journal}{Modelling Simul. Matter. Sci. Eng.}
  \textbf{\bibinfo{volume}{13}}, \bibinfo{pages}{683} (\bibinfo{year}{2005}).

\bibitem[{\citenamefont{Cox et~al.}(1997)\citenamefont{Cox, Johnston, and
  Murrell}}]{mm}
\bibinfo{author}{\bibfnamefont{H.}~\bibnamefont{Cox}},
  \bibinfo{author}{\bibfnamefont{R.~L.} \bibnamefont{Johnston}},
  \bibnamefont{and} \bibinfo{author}{\bibfnamefont{J.~N.}
  \bibnamefont{Murrell}}, \bibinfo{journal}{Surf. Sci.}
  \textbf{\bibinfo{volume}{373}}, \bibinfo{pages}{67} (\bibinfo{year}{1997}).

\bibitem[{\citenamefont{Voter and Chen}(1987)}]{vc}
\bibinfo{author}{\bibfnamefont{A.~F.} \bibnamefont{Voter}} \bibnamefont{and}
  \bibinfo{author}{\bibfnamefont{S.~P.} \bibnamefont{Chen}},
  \bibinfo{journal}{Mater. Res. Soc. Symp. Proc.}
  \textbf{\bibinfo{volume}{82}}, \bibinfo{pages}{175} (\bibinfo{year}{1987}).

\bibitem[{\citenamefont{Gupta}(1981)}]{gupta}
\bibinfo{author}{\bibfnamefont{R.~P.} \bibnamefont{Gupta}},
  \bibinfo{journal}{Phys. Rev. B} \textbf{\bibinfo{volume}{23}},
  \bibinfo{pages}{6265} (\bibinfo{year}{1981}).

\bibitem[{\citenamefont{Sutton and Chen}(1990)}]{sc}
\bibinfo{author}{\bibfnamefont{A.~P.} \bibnamefont{Sutton}} \bibnamefont{and}
  \bibinfo{author}{\bibfnamefont{J.}~\bibnamefont{Chen}},
  \bibinfo{journal}{Philos. Mag. Lett.} \textbf{\bibinfo{volume}{61}},
  \bibinfo{pages}{139} (\bibinfo{year}{1990}).

\bibitem[{\citenamefont{Ercolessi and Adams}(1994)}]{glue}
\bibinfo{author}{\bibfnamefont{F.}~\bibnamefont{Ercolessi}} \bibnamefont{and}
  \bibinfo{author}{\bibfnamefont{J.~B.} \bibnamefont{Adams}},
  \bibinfo{journal}{Europhys. Lett.} \textbf{\bibinfo{volume}{26}},
  \bibinfo{pages}{583} (\bibinfo{year}{1994}).

\bibitem[{\citenamefont{Doye}({\natexlab{a}})}]{jon}
\bibinfo{author}{\bibfnamefont{J.~P.~K.} \bibnamefont{Doye}},
  \bibinfo{note}{unpublished}.

\bibitem[{not()}]{note1}
\bibinfo{note}{As the $n/m$ value for the Al Sutton-Chen potential (7/6) is
  close to one, in the effective pair format, there is only a shallow well in
  the pair potential and so the observed structures are somewhat similar to
  those seen for Zn and Cd Gupta potentials, which also have very shallow wells
  in their effective pair potentials\cite{jon2}}.

\bibitem[{\citenamefont{Haberland et~al.}(2005)\citenamefont{Haberland,
  Hippler, Donges, Kostko, Schmidt, and von Issendorff}}]{haberland}
\bibinfo{author}{\bibfnamefont{H.}~\bibnamefont{Haberland}},
  \bibinfo{author}{\bibfnamefont{T.}~\bibnamefont{Hippler}},
  \bibinfo{author}{\bibfnamefont{J.}~\bibnamefont{Donges}},
  \bibinfo{author}{\bibfnamefont{O.}~\bibnamefont{Kostko}},
  \bibinfo{author}{\bibfnamefont{M.}~\bibnamefont{Schmidt}}, \bibnamefont{and}
  \bibinfo{author}{\bibfnamefont{B.}~\bibnamefont{von Issendorff}},
  \bibinfo{journal}{Phys. Rev. Lett.} \textbf{\bibinfo{volume}{94}},
  \bibinfo{pages}{035701} (\bibinfo{year}{2005}).

\bibitem[{\citenamefont{Doye and Wales}(1997)}]{morse}
\bibinfo{author}{\bibfnamefont{J.~P.~K.} \bibnamefont{Doye}} \bibnamefont{and}
  \bibinfo{author}{\bibfnamefont{D.~J.} \bibnamefont{Wales}},
  \bibinfo{journal}{J. Chem. Soc., Faraday Trans.}
  \textbf{\bibinfo{volume}{93}}, \bibinfo{pages}{4233} (\bibinfo{year}{1997}).

\bibitem[{\citenamefont{Doye and Meyer}(2005)}]{blj}
\bibinfo{author}{\bibfnamefont{J.~P.~K.} \bibnamefont{Doye}} \bibnamefont{and}
  \bibinfo{author}{\bibfnamefont{L.}~\bibnamefont{Meyer}},
  \bibinfo{journal}{Phys. Rev. Lett.} \textbf{\bibinfo{volume}{95}},
  \bibinfo{pages}{063401} (\bibinfo{year}{2005}).

\bibitem[{\citenamefont{Frenkel and Smit}(2002)}]{frenkel}
\bibinfo{author}{\bibfnamefont{D.}~\bibnamefont{Frenkel}} \bibnamefont{and}
  \bibinfo{author}{\bibfnamefont{B.}~\bibnamefont{Smit}},
  \emph{\bibinfo{title}{Understanding Molecular Simulation}}
  (\bibinfo{publisher}{Academic Press, New York}, \bibinfo{year}{2002}).

\bibitem[{\citenamefont{Wales and Doye}(1997)}]{wales}
\bibinfo{author}{\bibfnamefont{D.~J.} \bibnamefont{Wales}} \bibnamefont{and}
  \bibinfo{author}{\bibfnamefont{J.~P.~K.} \bibnamefont{Doye}},
  \bibinfo{journal}{J. Phys. Chem. A} \textbf{\bibinfo{volume}{101}},
  \bibinfo{pages}{5111} (\bibinfo{year}{1997}).

\bibitem[{\citenamefont{Johnson and Oh}(1989)}]{johnson}
\bibinfo{author}{\bibfnamefont{R.~A.} \bibnamefont{Johnson}} \bibnamefont{and}
  \bibinfo{author}{\bibfnamefont{D.~J.} \bibnamefont{Oh}}, \bibinfo{journal}{J.
  Mater. Res.} \textbf{\bibinfo{volume}{4}}, \bibinfo{pages}{1195}
  (\bibinfo{year}{1989}).

\bibitem[{\citenamefont{Cleri and Rosato}(1993)}]{cleri}
\bibinfo{author}{\bibfnamefont{F.}~\bibnamefont{Cleri}} \bibnamefont{and}
  \bibinfo{author}{\bibfnamefont{V.}~\bibnamefont{Rosato}},
  \bibinfo{journal}{Phys. Rev. B} \textbf{\bibinfo{volume}{48}},
  \bibinfo{pages}{22} (\bibinfo{year}{1993}).

\bibitem[{\citenamefont{Sandberg et~al.}(2002)\citenamefont{Sandberg,
  Magyari-K{\"o}pe, and Mattsson}}]{sandberg}
\bibinfo{author}{\bibfnamefont{N.}~\bibnamefont{Sandberg}},
  \bibinfo{author}{\bibfnamefont{B.}~\bibnamefont{Magyari-K{\"o}pe}},
  \bibnamefont{and} \bibinfo{author}{\bibfnamefont{T.~R.}
  \bibnamefont{Mattsson}}, \bibinfo{journal}{Phys. Rev. Lett.}
  \textbf{\bibinfo{volume}{89}}, \bibinfo{pages}{065901}
  (\bibinfo{year}{2002}).

\bibitem[{\citenamefont{Jellinek and Goldberg}(2000)}]{jellinek}
\bibinfo{author}{\bibfnamefont{J.}~\bibnamefont{Jellinek}} \bibnamefont{and}
  \bibinfo{author}{\bibfnamefont{A.}~\bibnamefont{Goldberg}},
  \bibinfo{journal}{J. Chem. Phys.} \textbf{\bibinfo{volume}{113}},
  \bibinfo{pages}{2570} (\bibinfo{year}{2000}).

\bibitem[{\citenamefont{Werner}(2005)}]{werner}
\bibinfo{author}{\bibfnamefont{R.}~\bibnamefont{Werner}},
  \bibinfo{journal}{Eur. Phys. J. B} \textbf{\bibinfo{volume}{43}},
  \bibinfo{pages}{47} (\bibinfo{year}{2005}).

\bibitem[{\citenamefont{Sun and Gong}(1998)}]{sun}
\bibinfo{author}{\bibfnamefont{D.~Y.} \bibnamefont{Sun}} \bibnamefont{and}
  \bibinfo{author}{\bibfnamefont{X.~G.} \bibnamefont{Gong}},
  \bibinfo{journal}{Phys. Rev. B} \textbf{\bibinfo{volume}{57}},
  \bibinfo{pages}{4730} (\bibinfo{year}{1998}).

\bibitem[{\citenamefont{Doye}({\natexlab{b}})}]{lead}
\bibinfo{author}{\bibfnamefont{J.~P.~K.} \bibnamefont{Doye}},
  \bibinfo{journal}{Comp. Mater. Sci.} \bibinfo{pages}{in press}.

\bibitem[{\citenamefont{Nelson and Spaepen}(1989)}]{nelson}
\bibinfo{author}{\bibfnamefont{D.~R.} \bibnamefont{Nelson}} \bibnamefont{and}
  \bibinfo{author}{\bibfnamefont{F.}~\bibnamefont{Spaepen}},
  \bibinfo{journal}{Solid State Phys.} \textbf{\bibinfo{volume}{42}},
  \bibinfo{pages}{1} (\bibinfo{year}{1989}).

\bibitem[{\citenamefont{Frantz}(2001)}]{frantz}
\bibinfo{author}{\bibfnamefont{D.~D.} \bibnamefont{Frantz}},
  \bibinfo{journal}{J. Chem. Phys.} \textbf{\bibinfo{volume}{115}},
  \bibinfo{pages}{6136} (\bibinfo{year}{2001}).

\bibitem[{\citenamefont{Schmidt et~al.}(2003)\citenamefont{Schmidt, Donges,
  Hippler, and Haberland}}]{schmidt}
\bibinfo{author}{\bibfnamefont{M.}~\bibnamefont{Schmidt}},
  \bibinfo{author}{\bibfnamefont{J.}~\bibnamefont{Donges}},
  \bibinfo{author}{\bibfnamefont{T.}~\bibnamefont{Hippler}}, \bibnamefont{and}
  \bibinfo{author}{\bibfnamefont{H.}~\bibnamefont{Haberland}},
  \bibinfo{journal}{Phys. Rev. Lett.} \textbf{\bibinfo{volume}{90}},
  \bibinfo{pages}{103401} (\bibinfo{year}{2003}).

\bibitem[{\citenamefont{Calvo and Spiegelmann}(2000)}]{calvo}
\bibinfo{author}{\bibfnamefont{F.}~\bibnamefont{Calvo}} \bibnamefont{and}
  \bibinfo{author}{\bibfnamefont{F.}~\bibnamefont{Spiegelmann}},
  \bibinfo{journal}{J. Chem. Phys.} \textbf{\bibinfo{volume}{112}},
  \bibinfo{pages}{2888} (\bibinfo{year}{2000}).

\bibitem[{\citenamefont{Doye and Calvo}(2002)}]{doye-4}
\bibinfo{author}{\bibfnamefont{J.~P.~K.} \bibnamefont{Doye}} \bibnamefont{and}
  \bibinfo{author}{\bibfnamefont{F.}~\bibnamefont{Calvo}}, \bibinfo{journal}{J.
  Chem. Phys.} \textbf{\bibinfo{volume}{116}}, \bibinfo{pages}{8307}
  (\bibinfo{year}{2002}).

\bibitem[{\citenamefont{Doye}(2003{\natexlab{b}})}]{jon2}
\bibinfo{author}{\bibfnamefont{J.~P.~K.} \bibnamefont{Doye}},
  \bibinfo{journal}{Phys. Rev. B} \textbf{\bibinfo{volume}{68}},
  \bibinfo{pages}{195418} (\bibinfo{year}{2003}{\natexlab{b}}).

\end{thebibliography}

%%%%%%%%%%%%%%%%%%%%%%%%%%%%%%%%%%%%%%%%%%%%%%%%%%%%%%%%%%%%%%%%%%%%%%
%%%%%%%%%%%%%%%%%%%%%%%% LIST OF FIGURES %%%%%%%%%%%%%%%%%%%%%%%%%%%%%%

%\begin{widetext}

%
%\begin{figure}
%\begin{center}
%\includegraphics[width=170mm]{figure2}
%\caption{\label{fig2} }
%\end{center}
%\end{figure}
%
%\begin{figure}
%\begin{center}
%\includegraphics[width=170mm]{figure3}
%\caption{\label{fig3} }
%\end{center}
%\end{figure}
%
%\end{widetext}
%
\end{document}